\documentclass[a4paper]{article}

\usepackage{amsmath,amssymb}

\newcommand{\ii}{\mathrm{i}}

\newcommand{\pd}{\partial}

\newcommand{\e}{\mathrm{e}}

\newcommand{\q}{\tilde{\theta}}
\newcommand{\sgn}{\mathop{\mathrm{sgn}}}

\begin{document}

\title{Can string bits be supersymmetric?}
\author{Stefano Bellucci\thanks{e-mail:
\texttt{bellucci@lnf.infn.it}}, Corneliu Sochichiu\thanks{On
leave from: \textit{Bogoliubov Laboratory of Theoretical Physics,
Joint Institute for Nuclear Research, 141980 Dubna, Moscow Reg.,
RUSSIA}; e-mail: \texttt{sochichi@lnf.infn.it}}
\\
\\
{\it INFN-Laboratori Nazionali di Frascati,}\\
{\it Via E. Fermi 40, 00044 Frascati, Italy}}

\maketitle
\begin{abstract}
We search for the possibility to have supersymmetric string
bits at finite discretization $J$. From a general setup we find
that the string bits can be made supersymmetric modulo a single
defect mode which is not expected to have any sensible effect
in the continuum limit.
\end{abstract}

\section{Introduction}
IIB string in plane wave background
\cite{Metsaev:2001bj,Metsaev:2002re} has drawn a considerable
interest due to its relation to AdS/CFT correspondence (see
\cite{Aharony:1999ti} for a comprehensive review and
\cite{Bianchi:2003wx} for recent progress). It arises as a limit
(Penrose limit) of AdS geometry
\cite{Blau:2001ne,Blau:2002dy,Blau:2002mw}. This corresponds to
the limit in AdS/CFT correspondence when one considers
Super--Yang--Mills (SYM) operators with large supersymmetry
$R$-charge $J$ \cite{Berenstein:2002jq,Berenstein:2002zw}. It
appears that in this limit one can extend the perturbative
analysis of the SYM model to large values of 't Hooft couplings.
The latter allows one to have a reliable perturbation knowledge of
both sides of the correspondence.

The string bit model
\cite{Verlinde:2002ig,Vaman:2002ka,Pearson:2002zs} was proposed as
a useful tool to describe the stringy plane wave dynamics at
finite $J$. In this model the continuum string is replaced by a
finite number of elastically interacting points
--- string bits. (This number is identified with the $R$-charge
$J$ in SYM model.) Although it succeeded to produce a good
agreement with the SYM results in the bosonic sector, it appeared to
contain inconsistencies \cite{Bellucci:2003qi}, due to a
spectrum doubling problem in the fermionic sector (see e.g.
\cite{Creutz} for details related to fermion doubling).

Although the problems related to doubling was discussed earlier in
the context of flat space superstrings in \cite{Bergman:1995wh},
the existence of a consistent discretization of the pp-wave string is
particularly important, as it justifies the consistency of the
correspondence at finite values of the $R$-charge $J$, as well as
its limit $J\to\infty$.

A way to cure the fermion doubling by a modification in the
fermionic sector of the model was proposed recently in
\cite{Danielsson:2003yc}. For this the \emph{staggered}, or
\emph{Kogut--Susskind} fermion approach \cite{Kogut:1975ag} was
used. The idea of the method consists in the reduction of the
total number of discrete fermions by placing fermions of different
chiralities at different lattice sites, in such a way that the
fermion doubling at the end produces the right spectrum. The
drawback of this approach is that it violates supersymmetry at
finite values of lattice discretization. On the other hand, the finite
$J$ sector of Yang--Mills theory is explicitly supersymmetric, hence
one would expect that it should be described by a supersymmetric
effective discrete model of superstrings, even at finite charge
$J$. This would imply the existence of a \emph{supersymmetric}
string bit model at finite discretization parameter. The aim of the
present work is to study the ``natural'' limits in existence of
such a model.

The plan of the paper is as follows. First, we very briefly
introduce the string bit model and the fermion doubling problem.
Then, under quite generic assumptions we analyze the possible
forms of the fermionic part of the Hamiltonian, which may lead to a
supersymmetric model at finite $J$ having the correct fermionic
spectrum in the continuum limit.

\section{String bits and doublers}
The naive string bit model after fixing the permutation symmetry
is given in the one-string sector by the Hamiltonian
\begin{multline}\label{hamiltonian}
  H_{naive}=\sum_{n=0}^{J-1}\left[\frac{a}{2}(p_{in}^2+x_{in}^2)
  +\frac{1}{2a}(x^i_{n+1}-x^i_n)^2\right]\\
  -\frac{\ii}{2} \sum_{n=0}^{J-1}\left[(\theta_n\theta_{n+1}-
  \q_n\q_{n+1})-2a\q_n\Pi\theta_n\right],
\end{multline}
and commutation relations
\begin{equation}\label{PB}
  [p^i_n,x^i_n]=-\frac{\ii}{a}\delta^{ij}\delta_{mn},\quad
  \{\theta^a_n,\theta^b_m\}=\frac{1}{a}\delta^{ab}\delta_{mn},\quad
  \{\q^a_n,\q^b_m\}=\frac{1}{a}\delta^{ab}\delta_{mn},
\end{equation}
where $i=1,\dots,8$ are vector and, respectively, $a=1,\dots,8$
spinor indices of $SO(8)$ appearing in the light-cone quantization
of the pp-wave string. $\Pi$ is the matrix in $SO(8)$ spinor space
given in terms of $16\times 16$ dimensional $\gamma$-matrices in
chiral representation by\footnote{We abide by the notations of
\cite{Metsaev:2001bj}.}
\begin{equation}\label{Pi}
    \Pi=\gamma^1\bar{\gamma}^2\gamma^3\bar{\gamma}^4,\qquad \Pi^2=1.
\end{equation}

As we have shown in \cite{Bellucci:2003qi}, the model
\eqref{hamiltonian}, \eqref{PB} suffers from fermion spectrum
doubling. The doubling is most easily seen if one rewrites the
Hamiltonian in terms of Fourier modes  $x_k$, $p_k$ $\theta_k$ and
$\q_k$
\begin{equation}
  f_{n}=\frac{1}{\sqrt{J}}\sum_{k=-J/2}^{J/2-1}f_k\e^{2\pi\ii
  kn/J},
\end{equation}
where $f_n$ stands for $x_n$, $p_n$ $\theta_n$ and $\q_n$ while
$f_k$ represents their Fourier modes. Then the fermionic part of
the Hamiltonian takes the form,
\begin{equation}\label{H_f}
    H_f=\frac{1}{2}\sum_k\left[
    \sin\frac{2\pi k}{J}(\theta_{-k}\theta_{k}-\q_{-k}\q_{k})-2a\ii
    \q_{-k}\Pi\theta_k\right].
\end{equation}
>From eq. \eqref{H_f} one can see that, due to an extra zero of the sin
function at $k=J/2$, there are additional propagating modes of the
momenta at the ``edge of the Brillouin zone'' $|k|\sim J/2$.
Although the wavelengths of such modes are of the order of lattice size,
which corresponds to discontinuous fields in the continuum limit
$J\to\infty$, they fail to decouple since the energy they carry is
as small as that of ``good'' modes, which correspond to continuous
functions.

In fact, in analogy to the flat space case studied in
\cite{Bergman:1995wh}, the unwanted modes can be removed by the so
called Wilson term, i.e. the following second derivative term:
\begin{equation}\label{wilson-t}
    \Delta H_{W}=-\ii\sum_k\left[(\q_{n+1}-\q_n)\Pi
    (\theta_{n+1}-\theta_{n})\right].
\end{equation}
For slowly oscillating modes this produces a correction of the
order of the discretization spacing $a$, while for doubling modes it
produces a mass of the order of $1/a$ and therefore it makes them decouple,
in the continuum limit. This approach can be seen as an alternative to that
of Ref. \cite{Danielsson:2003yc}, and as it is not difficult
to see that it suffers from the same defect: at finite $J$,
supersymmetry is broken.

\section{Seeking supersymmetry}
Let us fix the general setup under which we analyze the
possibilities to build a supersymmetric Hamiltonian. In fact we
will act in a more or less straightforward way: we choose the
original bosonic Hamiltonian and try to find its fermionic
extension through probing the definitions of the supercharges.

In momentum representation the general Ansatz for the Hamiltonian
is chosen to be
\begin{multline}\label{ansatz}
  H=\frac{1}{2}\sum_k\left[ a\left(|p_{ik}|^2+
  (\tilde{\pd}_{-k}\tilde{\pd}_{k}+1)|x^i_{k}|^2\right)\right.\\
  -\ii \left.\left((1/2)(\hat{\pd}_{-k}-\hat{\pd}_{k})(\theta_{-k}\theta_{k}-
  \q_{-k}\q_{k})-2a\q_{-k}\Pi\theta_{k}\right)\right],
\end{multline}
where the momentum function
\begin{equation}\label{derivative}
    \tilde{\pd}_{k}=(\e^{2\pi\ii k/J}-1)
\end{equation}
gives the bosonic derivative, while $\hat{\pd}_{k}$ is the
fermionic one to be found.

If one chooses the fermionic momentum function $\hat{\pd}_{k}$ to
be equal to the bosonic one $\hat{\pd}_{k}=\tilde{\pd}_{k}$ as in
the naive case, then, due to the fact that
$\hat{\pd}_{-k}=\hat{\pd}^*_{k}$, the energy spectrum of fermions
will depend on the imaginary part of the momentum function, in
contrast with the bosons, whose energy is given by the absolute value
of the momentum function. Generically, the imaginary part can have
more zeroes then the absolute value, which is what actually happens, thus
the extra zeroes are the source of doubling.

In order to find the fermionic Hamiltonian, let us adopt the following
strategy. Let us consider an Ansatz for the supercharges,
\begin{subequations}\label{q-ansatz}
  \begin{align}
    Q&=a\sum_k\left[p^i_{-k}\gamma^i\theta_{k}-
    x^i_{-k}\gamma^i\Pi\q_{k}+\bar{\pd}_{-k}x_{-k}\gamma^i\theta_{k}\right],\\
    \tilde{Q}&=a\sum_k\left[p^i_{-k}\gamma^i\q_{k}+
    x^i_{-k}\gamma^i\Pi\theta_{k}-\bar{\pd}_{-k}x_{-k}\gamma^i\q_{k}\right],
  \end{align}
\end{subequations}
where the unknown momentum function $\bar{\pd}_{k}$ appears. Of course, the
two unknown momentum functions $\bar{\pd}_{k}$ and $\hat{\pd}_{k}$
are not independent, they are related to each other as well as to
the bosonic momentum function through the would-be supersymmetry
algebra
\begin{equation}\label{susy}
    \{Q_a,Q_b\}=2\delta_{ab}(H+P), \quad
    \{\tilde{Q}_a,\tilde{Q}_b\}=2\delta_{ab}(H-P),\quad
    \{Q,\tilde{Q}\}=0.
\end{equation}

What remains to be done is just to check that the algebra \eqref{susy}
can be satisfied by an appropriate choice of $\bar{\pd}_{k}$ and
$\hat{\pd}_{k}$.

A direct computation of the (anti)commutators \eqref{susy}, using
the definition \eqref{q-ansatz} of the supercharges, yields
\begin{equation}\label{bar-susy}
    \{Q_a,Q_b\}=2\delta_{ab}(\bar{H}+\bar{P}), \qquad
    \{\tilde{Q}_a,\tilde{Q}_b\}=2\delta_{ab}(\bar{H}-\bar{P}).
\end{equation}
where
\begin{align}\nonumber
  \bar{H}&=\frac{a}{2}\sum_k
  \left[p^i_{-k}p^i_{k}+(1+\bar{\pd}_{-k}\bar{\pd}_{k})x^i_{-k}x^i_{k}+\right.\\
  \label{bar-H}
  &\qquad
  \left.\ii
  (\theta_{-k}\bar{\pd}_{k}\theta_{k}-\q_{-k}\bar{\pd}_{k}\q_{k})-
  2\ii \q_{-k}\Pi\theta_k\right],\\
  \label{bar-P}
  \bar{P}&=\frac{a}{2}\sum_k \left[2p^i_{-k}\bar{\pd}_{k}x_{k}+\ii
  (\theta_{-k}\bar{\pd}_{k}\theta_{k}+\q_{-k}\bar{\pd}_{k}\q_{k})\right],
\end{align}
while for the supercharge anticommutators one has
\begin{equation}\label{QQ}
    \{Q_{a},\tilde{Q}_{b}\}=a\sum_k(\bar{\pd}_{k}+\bar{\pd}_{-k})
    \left[(\gamma^i\Pi\gamma^j)_{ab} x^i_{-k}x^j_{k}-\ii
    \delta_{ab}\theta_{-k}\q_{k}\right].
\end{equation}

Obviously, one can identify the operators $\bar{H}$ and $\bar{P}$
with the Hamiltonian and shift operator respectively. In order to have
the correct algebra, one should also require the supercharge
commutator \eqref{QQ} to vanish. Identifying $\bar{H}$ with
the Hamiltonian requires that the fermionic momentum function
$\hat{\pd}$ coincide with $\bar{\pd}$, and that it should be related to
the bosonic momentum function through
\begin{equation}\label{bosonic-deriv}
    -\tilde{\pd}_{-k}\tilde{\pd}_{k}\equiv \tilde{\pd}^*_{k}\tilde{\pd}_{k}=
    \bar{\pd}_{-k}\bar{\pd}_{k}.
\end{equation}

On the other hand, the vanishing of the supercharge-supercharge
anticommutator requires the momentum function to be odd with
respect to the inversion of $k$,
\begin{equation}\label{imparity}
    (\bar{\pd}_{k}+\bar{\pd}_{-k})=0.
\end{equation}
Combining \eqref{bosonic-deriv} and \eqref{imparity} together, one
has
\begin{equation}\label{bar-to tilde}
    \bar{\pd}^2_{k}=-\tilde{\pd}^*_{k}\tilde{\pd}_{k}.
\end{equation}

Thus, the formal solution for $\bar{\pd}$ is given by the
operator square root
\begin{equation}\label{root}
    \bar{\pd}=\ii(\tilde{\pd}^*\tilde{\pd})^{\frac{1}{2}}.
\end{equation}

Obviously, in the continuum case the operator $\tilde{\pd}$ is just
a partial derivative, which is an anti-hermitian operator, and the
solution for $\bar{\pd}$ is $\tilde{\pd}$ itself. In contrast, in the
discrete case the next-to-neighbor derivative is not purely
anti-Hermitian and cannot solve the equation for both sides.

The equation \eqref{root} does not define the solution completely,
since it is ambiguous. One further needs to define the branch of
the square root as well. In the ``$k$-representation'' in which we
work the operators are diagonal, therefore the operator square
root extraction is reduced to the extraction of the square root of
each eigenvalue.\footnote{This holds only in the case of non-degenerate
eigenvalues.} Then, the problem of the ambiguity in the
operator square root is translated into the sign ambiguity
of each eigenvalue.

The natural way to fix the signs of the square root in accordance
with the condition \eqref{imparity}, and the requirements that the
momentum function of fermions is smooth and behaves like
$\bar{\pd}_k=\ii k+O(k^2 a)$ in the limit $J\to\infty$, is given by
the following choice:
\begin{equation}\label{signs}
    \bar{\pd}_{k}=\ii\sgn k \sqrt{\tilde{\pd}^*_{k}\tilde{\pd}_{k}}=
    \ii\sin \left(\frac{\pi k}{J}\right).
\end{equation}
In \eqref{signs} the radical sign denotes the positive root.
Passing back to the ``x-representation'' produces a non-local
operator with the kernel $\bar{\pd}_{mn}$
\begin{equation}\label{x-reps}
  \bar{\pd}_{nm}=\frac{2}{a}(-1)^{n-m}\frac{
    \cos (\frac{\pi }{2J})
    \sin (\frac{\pi(n-m) }{J})}
    {\cos (\frac{\pi }{J}) -
    \cos (\frac{2\pi (n-m)}{J})}.
\end{equation}
>From eq.\eqref{x-reps} one can see that for $|m-n|\ll J$ the
kernel decays as $\sim 1/(m-n)$, rather than having a
next-to-neighbor character.

Simple arguments from Morse theory show that this situation is
quite general. Aiming to the effects of discreteness the
derivative function $\bar{\pd}$ has to be periodic in $k$ with period
$J$,
\begin{equation}\label{periodicity}
    \bar{\pd}_{k+J}=\bar{\pd}_{k}.
\end{equation}
Hence, it follows that a periodic function satisfying the odd parity
condition \eqref{imparity} should have an odd number of either
additional zeroes or discontinuities.\footnote{Of course, Morse
theory applies to continuous functions. In our case we speak about
functions which at large $J$ can be interpolated with arbitrary
precision by such continuous functions.} In fact, the existence of
additional zeroes, needed for
satisfying the periodicity, is the reason of doubling in the case of
a naive discretization of fermions.

Like additional zeroes, discontinuities still create problems to
the physical consistency of the model in the continuum limit.
Thus, as it was shown in lattice QCD in the context of the so called SLAC
discretization \cite{Karsten:1979wh}, a momentum function
discontinuity produces an additional non-local and Lorentz
non-invariant contribution which, in particular, \emph{always}
cancel the chiral anomaly.

In spite of this, it was shown \cite{Slavnov:1998ef} that, by
introducing an additional regularization at a scale smaller and
correlated to the lattice cutoff, one can cure the effects of the
discontinuity at the Brillouin zone.

As a matter of fact, in the present case, it is a simple exercise to write down
a supersymmetry preserving Pauli--Villars regularization of pp-wave
string bits. Indeed, owing to the commutation of supersymmetries with the
Hamiltonian on the most of the Hilbert space, the straightforward
mode truncation $|k|<M\sim1/a^{1+\epsilon}$, $\epsilon>0$,
preserves supersymmetry and removes the doubling.

\section{Discussion}
In this paper we addressed the problem of supersymmetry in the
string bit model at a finite discretization $J$. We succeeded in
showing that supersymmetry can be preserved, avoiding at the same
time the doubling of fermions, in all modes of the
string bit except one. The latter can in no way be avoided, but it is an
isolated mode and produces no effect in the continuum limit.

We started from the bosonic part of the string bit model where the
string bit interactions include the nearest neighbors. The latter is
in accordance with the planarity property of the large $N$ gauge
theories. The fermionic action compatible with it appears not to
have this property of a next-to-neighbor character at finite $J$,
only the square of the Dirac operator has a next-to-neighbor character.
The most interesting objects depend on the square of the Dirac
operator, rather than on Dirac operator itself, which does not
violate the planarity.

Our analysis concerns the free string and, therefore, the
conclusions are valid and rigorous only for the free string. Although
the main interest for the application of the string bit model
is found in the interacting string case, where we can so far only
extrapolate our conclusions under certain conditions, the free
case is still important as it is leaving place for the
self-consistency of the BMN correspondence.

Of course, the detailed analysis of the case of interacting string
bits is still needed and it could not be ruled out completely that
avoiding doubling and keeping supersymmetry will not be
obstructed by interactions. (In this case, an interesting question
would be a ``susy-optimized'' fermionic action of the Gisparg--Wilson
type \cite{Ginsparg:1982bj}.)

The optimistic expectation, that most probably this is not the case,
is fuelled by two main arguments. The first one is that by a gauge choice
one can always make string bits free, even in the interacting string, and the
second one is that the supersymmetry violating defect can be isolated and
decoupled by a regularization procedure. (For the latter, however, it is
important to have always a supersymmetric regularization scheme at
hand.)
\subsection*{Acknowledgements}
We benefited from useful discussions with Niklas Beisert and
Konstantin Zarembo.

This work was partially supported by a RBRF grant \#02-01-00126
and support of leading scientific schools, the European
Community's Human Potential Programme under the contract
HPRN-CT-2000-00131 Quantum Spacetime, the INTAS-00-0254 \&
INTAS-00-0262 grants, the NATO Collaborative Linkage Grant
PST.CLG.979389 and the Iniziativa Specifica MI12 of the INFN
Commissione Nazionale IV.


\providecommand{\href}[2]{#2}\begingroup\raggedright\endgroup
\end{document}